# Optimization of photonic crystal nanocavities based on deep learning

Takashi Asano* and Susumu Noda
*Department of Electronic Science and Engineering, Kyoto University,
Kyoto 615-8510 Japan*
*Corresponding author: tasano@kuee.kyoto-u.ac.jp

**Abstract**

An approach to optimizing the *Q* factors of two-dimensional photonic crystal (2D-PC) nanocavities based on deep learning is proposed and demonstrated. We prepare a dataset consisting of 1000 nanocavities generated by randomly displacing the positions of many air holes of a base nanocavity and their *Q* factors calculated by a first-principle method. We train a four-layer neural network including a convolutional layer to recognize the relationship between the air holes' displacements and the *Q* factors using the prepared dataset. After the training, the neural network becomes able to estimate the *Q* factors from the air holes' displacements with an error of 13% in standard deviation. Crucially, the trained neural network can estimate the gradient of the *Q* factor with respect to the air holes' displacements very quickly based on back-propagation. A nanocavity structure with an extremely high *Q* factor of $1.58 \times 10^9$ is successfully obtained by optimizing the positions of 50 air holes over $\sim 10^6$ iterations, having taken advantage of the very fast evaluation of the gradient in high-dimensional parameter space. The obtained *Q* factor is more than one order of magnitude higher than that of the base cavity and more than twice that of the highest *Q* factors reported so far for cavities with similar modal volumes. This approach can optimize 2D-PC structures over a parameter space of a size unfeasibly large for previous optimization methods based solely on direct calculations. We believe this approach is also useful for improving other optical characteristics.

## 1. INTRODUCTION

Photonic nanocavities based on artificial defects in two-dimensional (2D) photonic crystal (PC) slabs have realized high quality (*Q*) factors from ~ thousand [1], tens of thousands [2], hundreds of thousands [3], millions [4-9], to more than ten million [10] together with small modal volumes ($V_{cav}$) of the order of one cubic wavelength or less. A higher *Q* factor increases the storage time of photons and light-matter interaction time, and a smaller $V_{cav}$ enhances the light matter interaction strength and decreases the footprint. There have been various efforts to increase the *Q* factors and/or *Q/V* of 2D-PC slab nanocavities both in theory and experiments [2–16]. The developed PC nanocavities have been used for various applications including ultracompact channel add/drop devices [1], nano-lasers [17], laser arrays for sensing [18], strongly coupled light-matter systems in solids [19,20], ultra-low-power consumption optical bi-stable systems [21], ultracompact and low-threshold all-Si Raman lasers [22], and photonic buffer memories [23-25]. However, further improvement is desirable for the realization of more advanced applications.

The fundamental design principle to increase the designed *Q* factor ($Q_{des}$) of 2D-PC nanocavities is well known: the wavevevtor components of the cavity electro-magnetic field within the light cone should be decreased as much as possible to reduce the radiation loss [11]. Many design methods including Gaussian envelope approaches [2,3], analytic inverse problem approaches [12,13], genetic algorithms [14], and leaky position visualization [15]. have been proposed to obtain a higher $Q_{des}$ while keeping a small $V_{cav}$. For example, a five-step heterostructure nanocavity comprising a defect waveguide with lattice constant modulation analytically designed to realize a Gaussian envelope function for the mode field was reported to have a $Q_{des}$ of $7\times10^8$ and a $V_{cav}$ of 1.3 cubic wavelength in the material $(\lambda/n)^3$ with an assistance of leaky mode visualization technique [12]. In addition, a two-step heterostructure nanocavity with a $Q_{des}$ of $1.4\times10^8$ and a $V_{cav}$ of 1.3 $(\lambda/n)^3$ was reported [9], where the positions of the eight air holes near the center of the cavity were tuned based on the leaky position visualization method [15]. Recently, the L4/3 cavity, in which positions of 22 air holes were tuned, was reported to have a $Q_{des}$ of $2.1\times10^7$ and a $V_{cav}$ of 0.32 $(\lambda/n)^3$ [16]. Although these approaches were successful, there still remain numbers of unused design freedoms in PC nanocavities, which are difficult to fully utilize due to the large cost to calculate the gradient of the $Q$ factor in the high dimensional structural parameter space in each step of structural optimization.

In this paper, we propose an approach to optimize 2D-PC nanocavities based on deep learning of the relationship between the nanocavities' structures and their $Q$ factors. We prepare a dataset consisting of 1000 different nanocavities whose air hole's positions are randomly but symmetrically displaced. Their $Q$ factors are calculated by a first-principle method where multiple parallel computation techniques can be fully utilized to reduce the computation time. Next, we train a four-layer artificial neural network (NN) including a convolutional layer using the dataset to recognize the relationship between the air hole displacement patterns and their $Q$ factors. The trained NN becomes able to predict the gradient of the $Q$ factor with respect to the air holes' displacements at a speed extremely faster than the first-principle calculation. This is used to optimize the displacements of many air holes (50) for large number of repetitions (>1,000,000). This optimization method demonstrates a very high $Q$ factor that exceeds one billion.

## 2. FRAMEWORK

In general, automatic structural optimization with respect to target characteristic value(s) requires at least three steps; (a) select a set of parameters that represents the structure to be optimized, (b) calculate the gradient of the target characteristic value(s) with respect to the structural parameters, and (c) modify the structural parameters based on the calculated gradient. (b) and (c) are repeated until the target value(s) saturates. In step (a), selecting all parameters that have a strong correlation with the target characteristic(s) is important. In step (b), fast evaluation of the gradient is required to ensure enough repetition of the optimization. However, it is difficult to fulfill these requirements when the structures to be optimized have large degrees of freedom and requires a large computation cost for the evaluation of the gradient.

This situation applies to $Q$ factor optimization in 2D-PC nanocavities, and we utilize a deep neural network (DNN) [26] to resolve these requirements. A DNN implements a complex non-linear function that associates a fixed-size input to a fixed size output through multiple units connected from layer to layer by linear and nonlinear operations. Because a DNN contains a large numbers of internal adjustable parameters (such as connection weights and biases) for tuning, it can approximate various input-output relationships once the internal parameters are properly tuned using many sets of example input-output data (training data). In particular, a DNN that contains convolutional layer(s), called a convolutional network (CNN), is very effective for learning the spatial features of input data [27]. Because such a CNN is effective for image processing [28], it is considered useful to learn the relationship between the structure of 2D-PC nanocavities and their $Q$ factors. Once we obtain a properly trained CNN using a dataset prepared by first-principle calculations, the gradient of the $Q$ factor with respect to the structural parameters can be estimated much faster than the direct calculations. Therefore, optimization of numbers of parameters can be repeated many times to fully exert the potential of 2D-PC structures, which is impossible by the method solely based on a direct calculation owing to the exponential increase of the computation cost with the increase of the dimension of the structural parameters. Our strategy is summarized as follows:

(I) Select a base cavity structure to be optimized.

(II) Select the type of structural parameter to be optimized. (e.g. air hole position, air hole size, air hole shape)

(III) Generate many 2D-PC nanocavities from the base structure by randomly fluctuating all structural parameters selected in (II) in an area much larger than the cavity field.

(IV) Calculate the Q factors of the nanocavities prepared in (I) by a first principle method, where many structures can be calculated separately in multiple parallel fashion to reduce the computation time.

(V) Prepare an NN to learn the relationship between the structural parameters and the Q factors.

(VI) Train the NN by the relationship between the subset of the parameters selected from (III) and the Q factors calculated in (IV).

(VII) Find the best set of parameters that minimizes the error between the Q factors predicted by the NN and those calculated by the first principle method.

(VIII) Starting from an initial cavity structure, gradually change the structural parameters selected in (VII) using the gradient of the Q factor with respect to the parameters predicted by the trained NN many times until the Q factor saturates.

(IX) Check the true Q factor of the optimized structure obtained in (VIII) by the first principle calculation.

## 3. RESULTS

In the followings, we describe the optimization of a two-step heterostructure nanocavity as an example. Figure 1 (a) shows the structure of the base nanocavity [step (I)], which is a two-step heterostructure nanocavity made of silicon slab with a thickness of 220 nm. The radii of air holes are 110 nm, and a line defect waveguide is formed by filling a row of air holes. The base lattice constant $a$ is 410 nm, and those around the center of the nanocavity are modulated by 3 nm in two steps as shown in the figure to confine light by the mode gap effect [3]. Eight air holes shown in the figure were shifted from their original positions by an order of $a/1000$ through a manual tuning process based on the leaky position visualization technique [15]. This manual tuning process increased the $Q$ factor of the nanocavity from 50 million (before tuning) to 140 million (after tuning) [10], but $V_{cav}$ was almost unchanged [~1.3 $(\lambda/n)^3$].

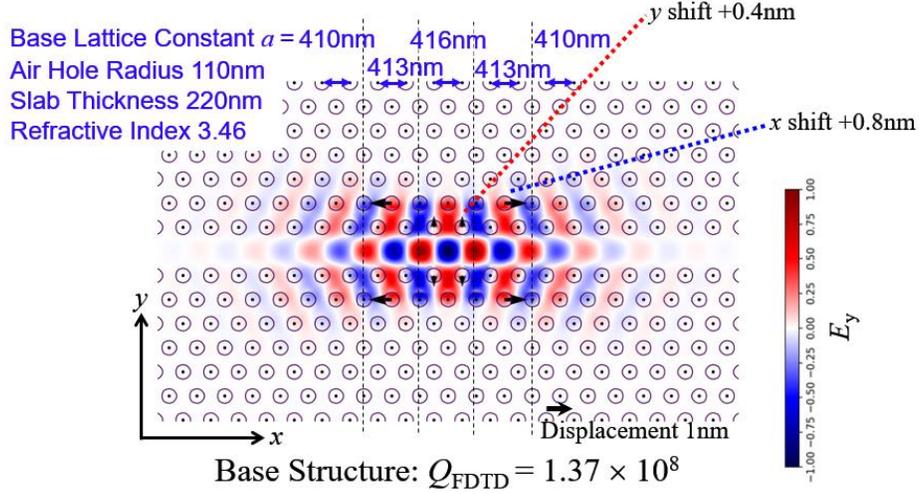

Fig. 1. Heterostructure two-dimensional photonic crystal nanocavity used as base structure for structural optimization. Circles indicate air holes with radii of 110nm formed in Si slab (refractive index $n = 3.46$) with thickness of 220 nm. Base lattice constant $a$ is 410 nm and lattice constants are modulated by 3 nm in $x$ direction in two-steps at center of nanocavity. Displacements of air holes from lattice positions are illustrated by vector arrows, where eight air holes are manually shifted by <1 nm. Electric field ($E_y$) distribution of fundamental resonant mode is plotted in color. Theoretical $Q$ factor of structure ($Q_{FDTD}$) is $1.37 \times 10^8$, and modal volume is 1.3 cubic wavelengths in material.

**A. Preparation of Dataset for Learning**

In step (II), we selected the displacements of the air holes as the parameters to be optimized. This is because the displacement of the air holes can be implemented in the fabrication process more accurately than other parameters such as the air hole radii or shapes. The air holes' positions can be accurately controlled by the electron beam writing process, while their radii and shapes are largely influenced by an etching process that is more difficult to control. In step (III), we added random displacements to all air holes in the $x$ and $y$ directions in such a way that the symmetry of the structure was maintained. We maintained the symmetry because an asymmetry of a PC cavity increases the radiation loss [11]. The induced random displacements obeyed a Gaussian distribution with a standard deviation of $a/1000$. This magnitude of the fluctuation was determined from the experience of the manual optimization mentioned above [10]. We generated 1000 randomly fluctuated nanocavity structures using procedure.

In step (IV), we calculated the electromagnetic field and the $Q$ factors of the fundamental resonant mode of the 1000 structures using the three-dimensional (3D) finite difference time domain (FDTD) method. A histogram of the calculated $Q$ factors ($Q_{FDTD}$) is plotted in Fig. 2, and examples of the generated cavity structures are shown in Fig. 3 with their electric fields ($E_y$) and $Q_{FDTD}$. It is seen in Fig. 2 that $Q_{FDTD}$ distributed in a range of almost two-orders of magnitudes, from $\sim 10^6$ to more than $3 \times 10^8$, but was mainly concentrated in the region below $\sim 3 \times 10^7$. In addition, there is a steep drop at the lower-$Q$ side of the peak. We thought that this nonuniform distribution of $Q_{FDTD}$ is relatively difficult to be learned by an NN [26], and transformed $Q_{FDTD}$ to $\log_{10}(Q_{FDTD})$ as shown in Fig. 2 (b), which shows a more uniform distribution similar to the Gaussian distribution. As a result, we used the relationships between the air hole's displacement patterns and $\log_{10}(Q_{FDTD})$s of the 1000 prepared structures as the dataset for the machine learning in step (VI).

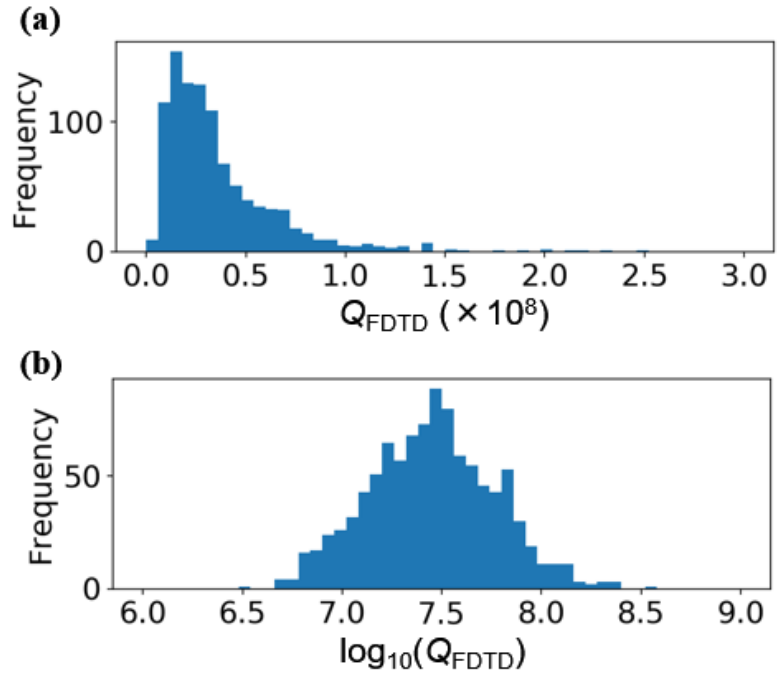

Fig. 2. (a) Histogram of $Q$ factors calculated by 3D-FDTD method ($Q_{FDTD}$) for 1000 nanocavities generated by randomly displacing positions of air holes of base shown in Fig. 1, where distribution of random displacement follows Gaussian distribution with standard deviation of 1/1000 lattice constant. (b) Histogram of $\log_{10}(Q_{FDTD})$.

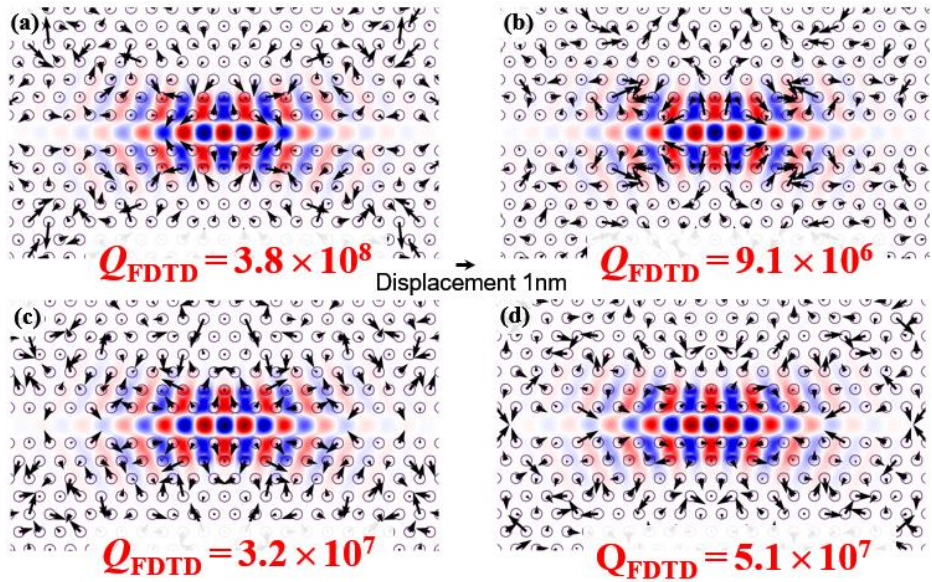

Fig. 3. (a)–(d) Examples of randomly generated cavity structures and their $Q$ factors calculated by 3D-FDTD method ($Q_{FDTD}$). Displacements of air holes are indicated by black arrows. Electric field ($E_y$) distribution of fundamental resonant mode is plotted in color.

## B. Configuration of Neural Network

Figure 4 shows the configuration of the neural network prepared in step (V). The input data were a set of displacement vectors $\vec{d}_{ij}$ of air holes [from the positions of a base cavity (I)] in an $N_x$ ($a$)× $N_y$ (rows) area around the center of the nanocavity that is normalized by the unit of $a/1000$, where $i$ and $j$ are the discrete $x$ and $y$ coordinate of the air holes. The first layer was a convolutional layer [27] in which $\vec{d}_{ij}$ in a local area of the input are summarized into one unit in the next layer by element-wise multiplication with a weight matrix of size $N_{fw}$ (holes)× $N_{fh}$ (rows) (called filter), and summation. By iteratively shifting the application area of this operation, where the amount of the shift is defined as stride, the input is convoluted with the filter to be summarized into a feature map. We used 50 filters of size $3 \times 5$ (×2channels: $x$ and $y$ displacements) so that the second layer contained 50 different summaries (feature maps) of the input, where the strides in the $x$ and $y$ directions were 1 and 2, respectively.

The second layer was fully connected to the third layer with 200 units through a rectified linear unit (ReLU [29]) and Affine transformation (multiplication with a weight matrix followed by summation with a bias vector). The third layer was fully connected to the fourth layer with 50 units through ReLU, random information selection units (Dropout [30]), and Affine transformation. Finally, the information in the fourth layer was summarized into the one output unit through ReLU and Affine transformation. This output unit was supposed to predict $\log_{10}(Q_{FDTD})$.

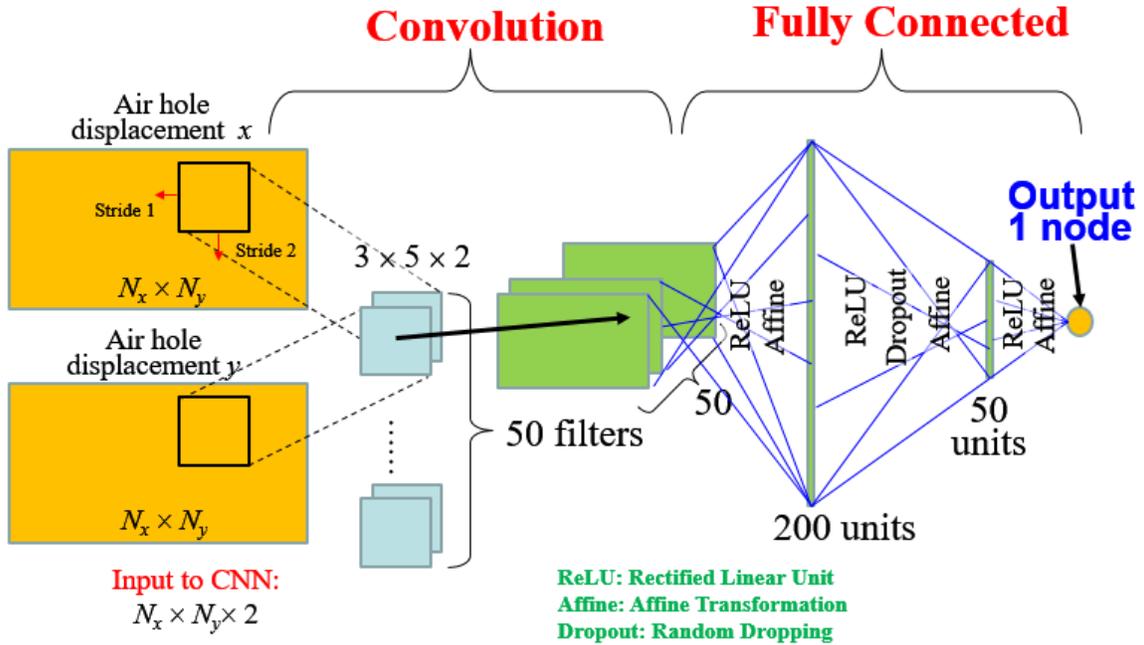

Fig. 4. Configuration of neural network prepared to learn relationship between displacements of air holes and $Q$ factors.

## C. Training of Neural Network

### 1. Loss function

In step (VI), we trained this neural network using 900 data among the 1000 prepared in steps (III and IV), and left the remaining 100 data as test datasets. A test dataset is necessary to avoid overfitting, which is a situation in which an NN is too optimized to the training examples so that it cannot predict meaningful answers for the other input. It is important to maximize the generalization ability of an NN, which is an ability to predict meaningful answers to new inputs it has never seen during the training. Therefore, the generalization ability of an NN should be checked by using a test dataset at intervals [26]. For the training, we set the loss function $L$ as

$$L = |Output - log_{10} Q_{FDTD}|^2 + \frac{1}{2}\lambda \sum_i w_i^2. \quad (1)$$

The first term of Eq. (1) represents the deviation from the true answer. The second term was introduced to penalize large connection weights $w_i$ in the network, where the summation is taken over all the weights in the network. This additional term is effective to prevent the overfitting (the weight decay method [31]), and $\lambda$ is the control parameter (we used $\lambda = 0.001$). We randomly selected one set of "$\vec{d}_{ij}$ pattern"-"$log_{10}(Q_{FDTD})$" from the training dataset, and gradually changed the internal parameters of the NN to reduce $L$ (stochastic gradient method [26]) based on the gradient of $L$ with respect to the internal parameters, which was obtained by the back-propagation method [32]. We also applied the Momentum optimization method to speed up the convergence [33], where the learning rate and the momentum decay rate were set as 0.001 and 0.9, respectively. The learning step was repeated until $L$ of the test dataset converges. The accuracy of the prediction was evaluated as the standard deviations of the output [$=log_{10}(Q_{NN})$] from the true answer [$log_{10}(Q_{FDTD})$] for the test dataset ($\sigma_{test}$) and training dataset ($\sigma_{train}$). These values were further converted into more comprehensive average prediction errors of the $Q$ values ($E_Q$) by the following equation:

$$E_{Qtest(train)} = (10^{\sigma_{test(train)}} - 10^{-\sigma_{test(train)}})/2 \quad (2)$$

This definition means that 68.27% of $Q_{NN}$'s are statistically distributed within $(1 \pm E_Q) \times Q_{FDTD}$'s

### 2. Example of training

An example of a learning curve [iteration of learning (optimization) v.s. prediction error] is shown in Fig. 5 (a), where input area size $(N_x, N_y)=(10, 5)$. It is seen in the figure that $E_Q$ for the training and test datasets are initially more than 80%, but decrease to less than 20% after ~2000 learning iterations. It is natural that $E_{Qtest}$ is always larger than $E_{Qtrain}$ because the NN has never learned the test dataset. Nevertheless, the minimum $E_{Qtest}$ became as small as ~16% within $2 \times 10^5$ iterations. The correlation between $Q_{FDTD}$ and $Q_{NN}$ for the training and test datasets (for the case with minimum $E_{Qtest}$) is shown in Fig. 5 (b) and (c), respectively. A good correlation with a correlation coefficient of 0.92 is obtained even for the test dataset. This result demonstrates successful achievement of the generalization ability, at least within the parameter space in which the prepared structure distributed (dimensions: $10 \times 5 \times 2$, range of each displacement: $\sim \pm a/1000$). We noticed that the correlation is better for the lower-$Q$ region, and the deviation from $Q_{FDTD}$ increases in the higher-$Q$ region. This is because the numbers of training datasets are much smaller in the higher-$Q$ region compared to the lower $Q$ region as shown in Fig. 2.

### 3. Dependence on input area size

Next, we trained the NN by changing the input area size $(N_x, N_y)$, and plotted the minimum $E_{Qtest}$ (obtained during 1 million iterations of the learning steps) as a function of $N_x$ and $N_y$ in Fig 6. It is seen in the figure that the prediction error is higher than 60% when $(N_x, N_y)$ is as small as (2, 5). The prediction error decreases as the input area size increases, and the case with $(N_x, N_y) = (13, 5)$ shows the minimum prediction error of ~13%, where the correlation coefficients between $Q_{FDTD}$ and $Q_{NN}$ for the test and training datasets were 0.96 and 0.99, respectively. However, the error increases again for a larger input area size. It is considered that the learning process was disturbed when the air holes' displacements that have small correlations with $Q_{FDTD}$ were input to the NN because they acted as noise for the learning process. This provides a hint: Fig. 6 allows us to pick up the structural parameters that have strong correlations with the target value, that is, the parameters that are most effective in the optimization process. In this case, we decided to optimize the displacement of airholes in areas with $(N_x, N_y) = (13, 5)$ in step (VII).

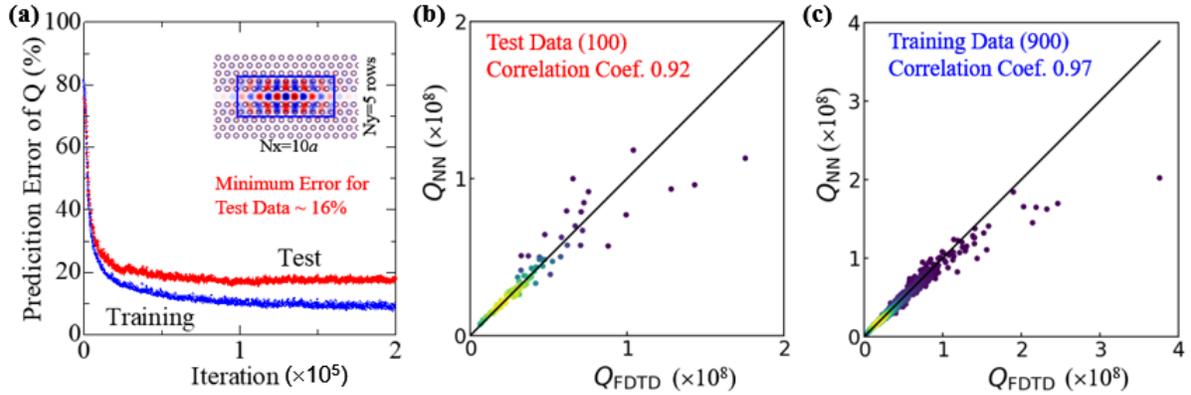

Fig. 5. Result of training for case where input air hole area $(N_x, N_y) = (10, 5)$. (a) Learning curve. (b), (c) Correlation between $Q_{FDTD}$ and prediction by neural network ($Q_{NN}$) for test and training dataset, respectively.

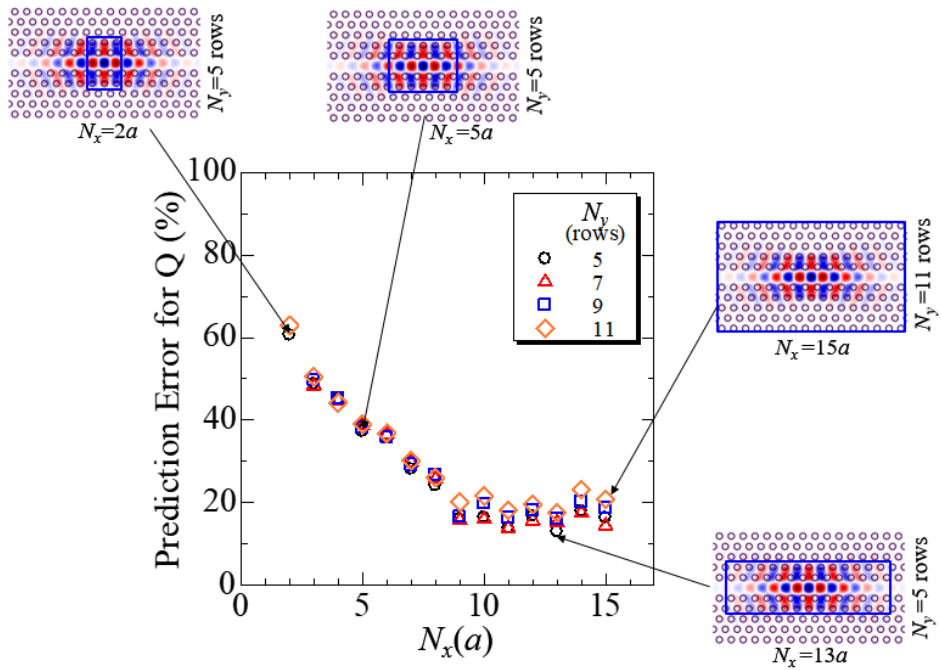

Fig. 6. Dependence of prediction error of $Q$ for test dataset on input air hole area size $(N_x, N_y)$.

## D. Structural Optimization by Trained Neural Network

### 1. Loss function

In step (VII), we performed a structural optimization of the nanocavity using the gradient method, where the gradient was calculated by the trained NN. More precisely, we took the advantage of the error back propagation method [32] to enable a high-speed calculation of the gradient. The back-propagation method is extremely effective to calculate the gradient of the loss with respect to the internal parameters of an NN, which were already used in the training process (VI). This method can be also used obtain the gradient of loss with respect to the input parameters (i.e., the air holes' displacements) rapidly. We set the loss $L'$ as $\left|\log_{10} Q_{NNW} - \log_{10} Q_{target}\right|^2$, and calculated the gradient of $L'$ with respect to $\vec{d}_{i,j}$ using the same framework used for the training process, where $Q_{target}$ was set to a very high constant value ($10^{10}$). We also added an artificial loss to penalize large displacements to constrain the air holes from moving far away from the parameter space that the NN learned in step (VI), where the prediction error was small:

$$L' = \left|log_{10} Q_{NNW} - log_{10} Q_{target}\right|^2 + \frac{1}{2}\lambda' \sum_{i,j} \left|\vec{d}_{i,j}\right|^2, \quad (3)$$

where $\lambda'$ is a control parameter. Then, we changed the structure ($\vec{d}_{i,j}$) step by step to reduce the loss $L'$ based on the Momentum [33] method:

$$\vec{d}_{i,j}^{n+1} = \vec{d}_{i,j}^{n} + \vec{v}_{i,j}^{n} \quad (4\text{-}1)$$

$$\vec{v}_{i,j}^{n+1} = \gamma \vec{v}_{i,j}^{n} - o_r \nabla L' \quad (4\text{-}2)$$

where $\vec{d}_{i,j}^{n}$ and $\vec{v}_{i,j}^{n}$ are the displacement and momentum of the air hole at position $i,j$ in the $n$-th step ($\vec{v}_{ij}^{0} = 0$), $\gamma$ is a control parameter called the momentum damping factor (=0.9), and $o_r$ is the optimization rate (=$1 \times 10^{-5}$). This process is iterated $10^6$ times.

### 2. Optimization curve

Figure 7 shows the evolution of $Q_{NN}$ during the optimization for various $\lambda$'s from 0.01 to 1, where the initial structure $\vec{d}_{i,j}^{0}$ was set to the structure that had the highest $Q_{FDTD}$ [=$3.8 \times 10^8$, Fig. 3 (a)] among the 1000 randomly prepared structures in step (III-IV). We also optimized the structure using $\lambda' = 0.001$, but the obtained result was identical to the case with $\lambda' = 0.01$. It is seen in the figure that $Q_{NN}$ increased from the original value after optimization in the cases with $\lambda' \leq 0.1$. $Q_{NN}$ slightly decreased from the original value after optimization in the cases with $\lambda' \geq 0.5$. This is considered because the high additional loss of $1/2\lambda'\left|\vec{d}_{i,j}\right|^2$ inhibited large deformation from the original structure (Fig. 1). Please note that the origin of $\vec{d}_{i,j}$ is the structure shown in Fig. 1, not the initial structure [Fig. 3 (a)]. We also carried out the optimization process with a completely different randomly created initial structure (denoted as fluc46398753) with $\lambda' = 0.05$. The result is plotted in Fig. 7 as indicated by the brown solid line. The initial $Q_{NN}$ of this structure is as low as $1 \times 10^7$, but increased to $4.84 \times 10^8$ after optimization. The final $Q_{NN}$ is almost the same as the case started from the structure in Fig. 3 (a) with $\lambda = 0.05$, and the obtained structures were also almost identical [see Fig. 8 (b) and (f)].

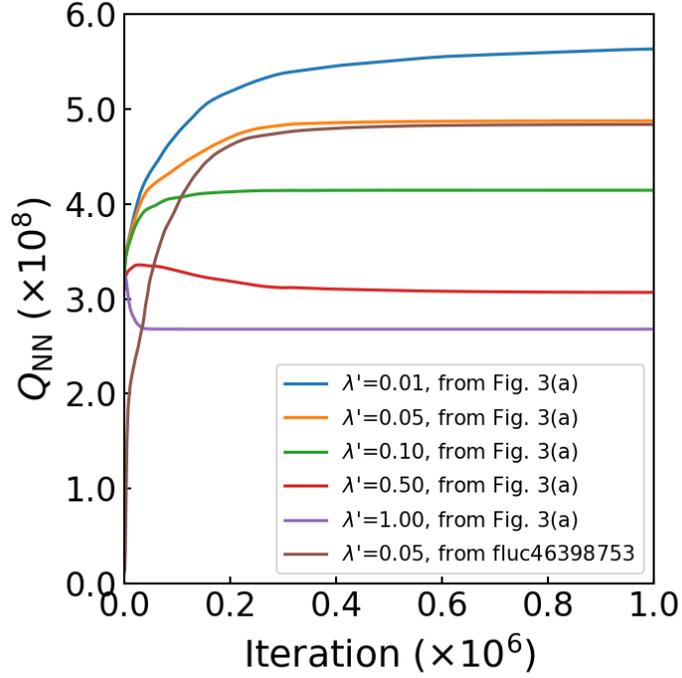

Fig. 7. Evolution of $Q_{NN}$ during the optimization process for different displacement constraint parameter $\lambda'$.

### 3. Validation by FDTD

In step (IX), we calculated the $Q$ factors of the structures obtained in step (VIII) by the 3D-FDTD method. The results are summarized in Fig. 8, where the structure after optimization, $Q_{FDTD}$, $Q_{NN}$, electric field distribution ($E_y$), and cavity modal volume ($V_{cav}$) are shown. It is seen in the figure that an extremely high $Q_{FDTD}$ of $1.58 \times 10^9$ was obtained with $\lambda' = 0.1$ [Fig. 8(c)]. This $Q_{FDTD}$ is one order of magnitude higher than the manually optimized two-step heterostructure nanocavity ($Q_{FDTD} = 1.37 \times 10^8$, Fig. 1), and more than twice of the highest $Q_{FDTD}$ of the 2D-PC nanocavity reported so far ($Q_{FDTD} = 7 \times 10^8$ [11]). The successful achievement of such an extremely high $Q_{FDTD}$ demonstrates the effectiveness of the proposed optimization method, where large degrees of freedom of the 2D-PC structure were effectively utilized, as can be seen from a comparison between Fig. 1 and Fig. 8.

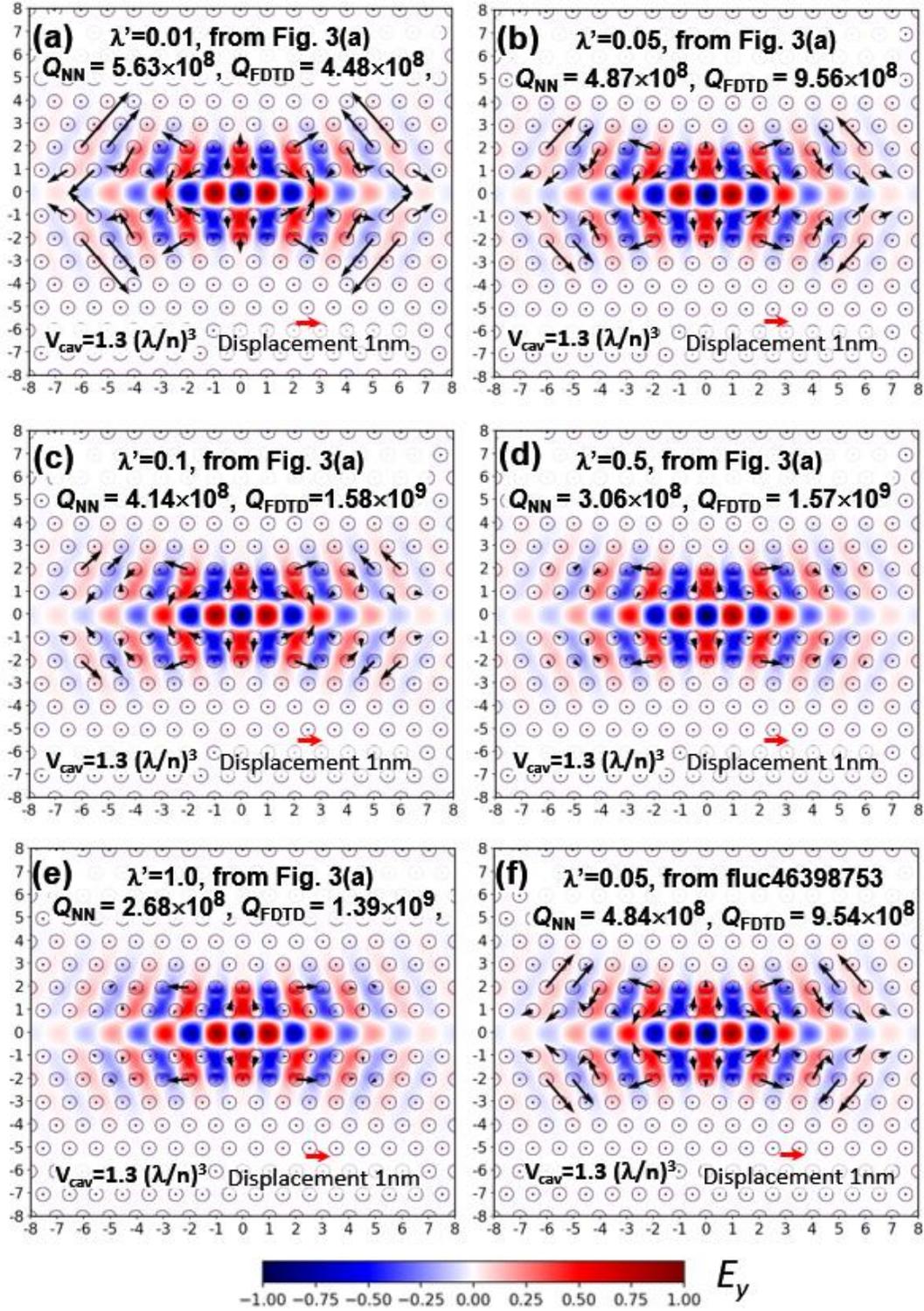

Fig. 8. Optimized structures and their $Q$ factors, electric field distributions ($E_y$), and modal volumes ($V_{cav}$).

## 4. DISCUSSION

It is seen in Fig. 8(c) that $Q_{NN}$ is less than 1/3 of $Q_{FDTD}$. This is understood from the response of the trained NN shown in Fig. 5 (b) and (c), where $Q_{NN}$ tends to be lower than $Q_{FDTD}$ as $Q_{FDTD}$ increases. As discussed before, the number of data with $Q_{FDTD} > 1 \times 10^8$ is rare (40 samples among 1000), and there are many data with lower $Q_{FDTD}$ (Fig. 2) so that the prediction tends to be low for high $Q_{FDTD}$ structures. Nevertheless, the fact that the structure optimized by this method has a larger $Q_{FDTD}$ than the initial structure indicates that the *direction* of the gradient of $Q_{FDTD}$ with respect to the air holes' displacements were properly evaluated by the trained NN.

It is also interesting that the highest $Q_{FDTD}$ was achieved by constraining the magnitude of the airholes' displacements to some extent by increasing $\lambda'$ in the loss function $L'$ [Eq. (3))]. In Fig. 8(a) to (c), $Q_{FDTD}$ increases from $4.48 \times 10^8$ to $1.58 \times 10^9$ as $\lambda'$ increases from 0.01 to 0.1, and the corresponding air holes' displacements decreases. The reason why the structure in Fig. 8(c) shows a much larger $Q_{FDTD}$ than that of Fig. 8(a) is because the accuracy of the $Q$ prediction becomes lower as displacements of the air holes move away from the center of the parameter space that the NN has learned (i.e. $\vec{d}_{i,j}$=0). For the case with $\lambda' > 0.5$, the magnitudes of the displacements are too constraint to obtain the highest $Q_{FDTD}$, but $Q_{FDTD}$'s > $1.39 \times 10^9$ were still realized.

Finally, we compare Fig. 8 (b) and (f). Although the initial structures for these two cases are completely different (not shown, but can be seen from the large difference between the initial $Q_{NN}$'s shown in Fig. 7), the final optimized structures and their $Q_{FDTD}$'s are almost the same. This result indicates that the structures obtained by this method are globally optimized at least within and near the parameter space that the NN learned.

## 5. CONCLUSION

We have proposed and demonstrated a novel approach to optimize 2D-PC nanocavities based on deep learning of the relationship between the nanocavities' structures and their $Q$ factors. We have successfully trained a neural network consisting of a convolutional layer and three fully connected layers using 1000 randomly generated nanocavities and their $Q$ factors. After the training, the convolutional neural network has become able to predict the $Q$ factors from the displacement patterns of air holes with an error of 13% in standard deviation. Structural optimization has been carried out by estimating the gradient of $Q$ with respect to the air holes' displacements using the trained neural network. A nanocavity structure with an extremely high theoretical $Q$ factor of $1.58 \times 10^9$, which is 10 times larger than that of the manually optimized base structure, and more than twice the highest $Q$ factor ever reported for 2D-PC cavities with similar modal volumes, has been successfully obtained. We attribute our unprecedentedly high $Q$ factor to the ability of our method to optimize the nanocavity over a parameter space of a size unfeasibly large for previous methods based solely on direct calculations. We believe this approach is effective for the optimization of various types of 2D-PC nanocavity structures, not only for increasing $Q$ factors but also for improving other target characteristics.

**Funding**. This work was partially supported by JSPS KAKENHI (15H03993) and the New Energy and Industrial Technology Development Organization (NEDO).

**Acknowledgment**. The authors would like to thank Mr. Koki Saito for his helpful textbook on deep learning written in Japanese (Deep learning from scratch, O'Reilly Japan).